\documentclass[12pt,preprint]{aastex} 
\shorttitle{Unusual CV KIC 9406652} 
\shortauthors{Gies et al.} 
 
\begin{document} 
 
\received{2013 June 10} 
\accepted{2013 August 1} 
 
\title{KIC 9406652: An Unusual Cataclysmic Variable in the Kepler Field of View\altaffilmark{1}}  
 
\author{Douglas R. Gies, Zhao Guo} 
\affil{Center for High Angular Resolution Astronomy and  
 Department of Physics and Astronomy,\\ 
 Georgia State University, P. O. Box 5060, Atlanta, GA 30302-5060, USA; \\ 
 gies@chara.gsu.edu, guo@chara.gsu.edu} 
 
\author{Steve B. Howell\altaffilmark{2}} 
\affil{NASA Ames Research Center, P.O. Box 1, M/S 244-30, Moffett Field, CA 94035, USA; \\ 
steve.b.howell@nasa.gov}

\author{Martin D. Still} 
\affil{NASA Ames Research Center, Moffett Field, CA 94035, USA; 
Bay Area Environmental Research Inst, Inc., 560 Third St., West Sonoma, CA 95476, USA; \\
martin.still@nasa.gov}

\author{Tabetha S. Boyajian} 
\affil{Department of Astronomy, Yale University, P. O. Box 208101, New Haven, CT 06520-8101, USA; \\
tabetha.boyajian@yale.edu}

\author{Abe J. Hoekstra, Kian J. Jek, Daryll LaCourse, Troy Winarski}
\affil{Planet Hunters Program, 
Department of Astronomy, Yale University, P. O. Box 208101, New Haven, CT 06520-8101, USA; \\
abejhoekstra@hotmail.com, kianjin@gmail.com, nhawkb@gmail.com, troywinarski@gmail.com}

\altaffiltext{1}{This work is based in part from the Planethunters.org Guest Observer program.}

\altaffiltext{2}{Visiting Astronomer, Kitt Peak National Observatory,
National Optical Astronomy Observatory, operated by the Association
of Universities for Research in Astronomy, Inc., under contract with
the National Science Foundation.}

\slugcomment{Accepted for ApJ; 08/01/2013} 
\paperid{91975}


\begin{abstract} 
KIC 9406652 is a remarkable variable star in the {\it Kepler}
field of view that shows both very rapid oscillations and 
long term outbursts in its light curve.   We present an 
analysis of the light curve over quarters 1 -- 15
and new spectroscopy that indicates that the object is a 
cataclysmic variable with an orbital period of 
6.108 hours.  However, an even stronger signal appears in the light 
curve periodogram for a shorter period of 5.753 hours, and 
we argue that this corresponds to the modulation of flux from 
the hot spot region in a tilted, precessing disk surrounding 
the white dwarf star.  We present a preliminary orbital 
solution from radial velocity measurements of features 
from the accretion disk and the photosphere of the companion. 
We use a Doppler tomography algorithm to reconstruct the 
disk and companion spectra, and we also consider how these
components contribute to the object's spectral energy distribution 
from ultraviolet to infrared wavelengths.  This target offers 
us a remarkable opportunity to investigate disk processes during 
the high mass transfer stage of evolution in cataclysmic variables. 
\end{abstract} 
 
 
 
\setcounter{footnote}{2} 
 
\section{Introduction}                              

Cataclysmic variable (CV) stars are evolved, interacting binary systems
in which mass transfer from a cool, Roche-filling, donor star
feeds a dynamic accretion disk surrounding a white dwarf, gainer
star (Warner 1995; Hellier 2001).  The physical processes of 
disk accretion can lead to flux variations over a wide variety of 
amplitudes and timescales (Honeycutt et al.\ 1998), and
long term photometric monitoring is key to our understanding 
of the mass accretion and loss processes.  The NASA {\it Kepler}
mission is now providing us with a particularly rich data 
set to explore these variations in detail.   Several dozen new 
CVs are now known in the {\it Kepler} FOV (Howell et al.\ 2013;
Scaringi et al.\ 2013), and detailed investigations 
from {\it Kepler} of previously known CVs are available
for V344~Lyr (Wood et al.\ 2011; Cannizzo et al.\ 2012), 
V447~Lyr (Ramsey et al.\ 2012), and V1504~Cyg (Cannizzo et al.\ 2012).

Here we report on the discovery of outbursts and fast variability in the
{\it Kepler} light curve of the star KIC~9406652 (= TYC~3556-325-1) that 
were identified through the work of the citizen - scientist Planet Hunters 
program\footnote{http://www.planethunters.org/} (Fischer et al.\ 2012). 
This is a relatively faint and blue star ($V=12.5$, $B-V=+0.1$; Everett et al.\ 2012), 
and its variable nature was first determined from quarter 1 data by Debosscher et al.\ (2011). 
The light curve shows evidence of both short period 
oscillations and quasi-monthly outbursts that bear some similarity 
to those observed in old novae and nova-like CVs (Honeycutt et al.\ 1998). 
We present a periodogram and wavelet analysis of the {\it Kepler} light curve
in section 2, and we discuss evidence for two key periodic signals 
in addition to the low frequency power related to the outbursts. 
In section 3, we present new time series spectra from 2013 April 
that we use to measure radial velocities and derive preliminary 
orbital elements.  We describe the nature of the spectra of the 
companion and accretion disk in section 4.  We discuss the rapid 
light curve variations in section 5, and we argue that they result from 
the changes in disk flux that occur as the gas stream strikes 
a tilted, precessing disk (Wood \& Burke 2007).

 
\section{Kepler Light Curve}                        
 
The flux of the target KIC~9406652 was recorded in long cadence mode in each 
observing quarter, and we obtained the Presearch Data Conditioning version
of the light curve from the data archive (almost identical to the 
Simple Aperture Photometry version of the light curve). 
The entire {\it Kepler} light curve from quarters 1 through 15 is shown 
in the lower panel of Figure~1 and a close up portion from quarter 7 is 
shown in Figure~2.  We see that that are outbursts on timescales of weeks
as well as rapid, smaller amplitude variations.  The recurrence 
times for the outbursts vary from 27 to 84 d, and they are often 
(although not always) characterized by slow rise followed immediately 
by a sharp decline or dip.  This is opposite to the fast rise and slower decline 
that is often observed in dwarf novae outbursts (Cannizzo et al.\ 2012).   
At peak outburst, the system is typically 
0.7 mag brighter than average, while the dip minima are often 0.8 mag 
fainter than average.  The outbursts usually have a duration of $\approx 7$~d, 
although a much longer event was recorded around BJD~2455240. 
The fast variation (with a cycle time of about 6 h) is seen almost all 
the time with a full amplitude of approximately 0.2~mag (or smaller 
at peak outburst).  

\begin{figure}
\begin{center} 
 {\includegraphics[angle=90,height=12cm]{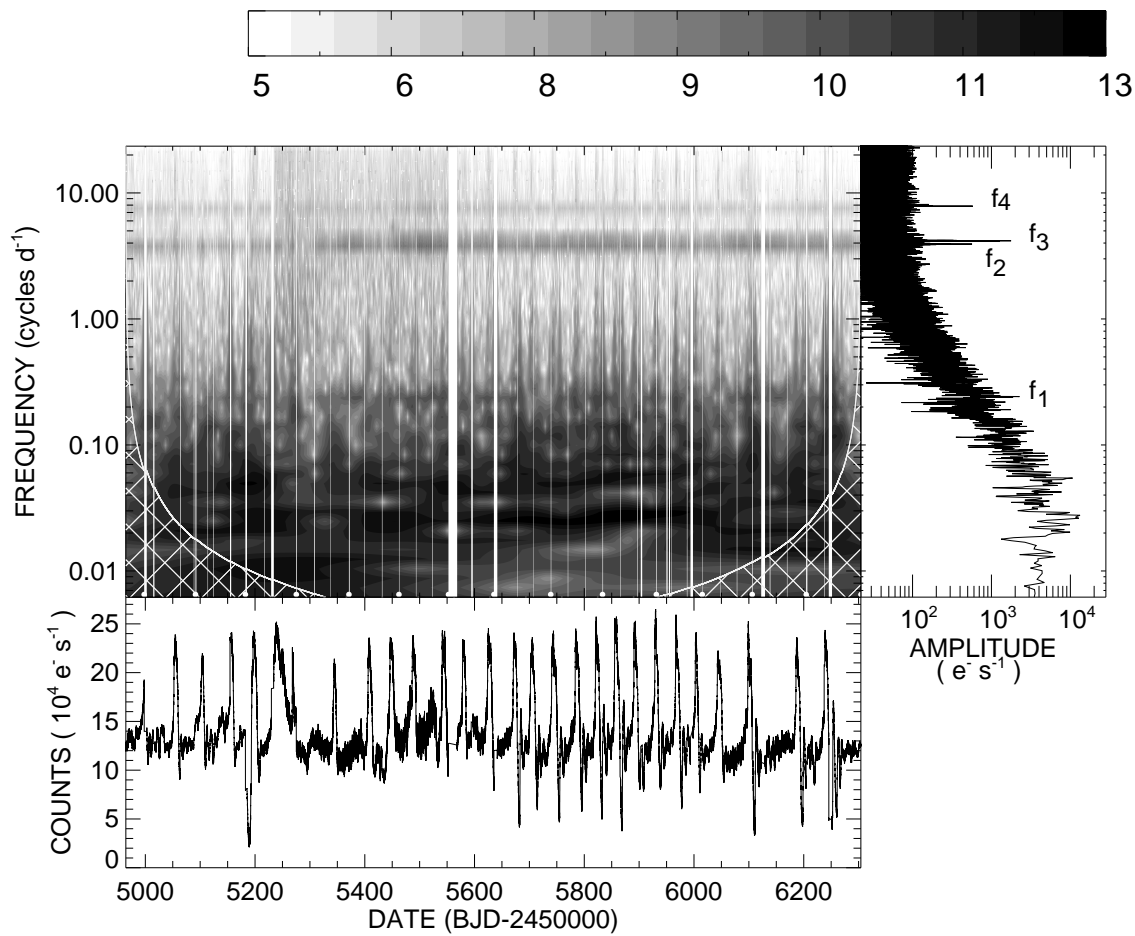}}
\end{center}
\caption{A grayscale image of the logarithm of the wavelet power as 
a function of time and frequency.  The white dots at the bottom of the 
image indicate the end of each quarter (1 -- 15), and the vertical 
white segments indicate gaps in the time series.  The cross-hatched 
regions in lower left and right show the ``cone of influence'' where
edge effects influence the wavelet power.  The upper legend shows the 
relation between gray intensity and logarithm (base 10) of the 
square of the wavelet amplitude.  The panel below shows the
corresponding {\it Kepler} light curve with time in units of 
Barycentric Julian Date.   The rotated panel to the right displays the 
amplitude of the full sample periodogram, and the four main signal 
frequencies are indicated to the right of the corresponding peaks. 
\label{fig1}} 
\end{figure}

\placefigure{fig1}     
 
\begin{figure} 
\begin{center} 
 {\includegraphics[angle=90,height=12cm]{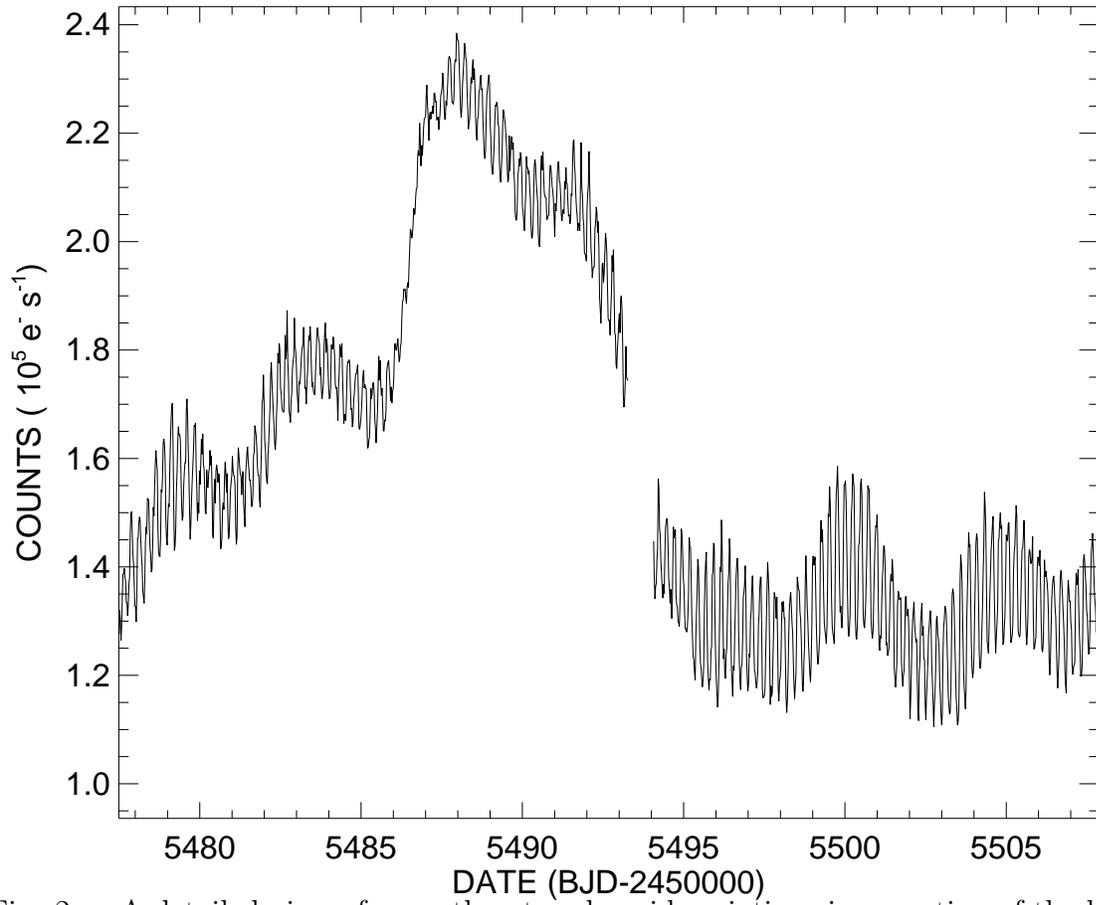}} 
\end{center} 
\caption{A detailed view of an outburst and rapid variations 
in a portion of the light curve from quarter 7. 
\label{fig2}} 
\end{figure}

\placefigure{fig2}     

The right hand panel in Figure~1 illustrates the periodogram for the entire
{\it Kepler} light curve from quarter 1 to 15.  There is a broad 
distribution of low frequency power that corresponds to the cyclic 
(but not strictly periodic) outbursts.  However, there are also four 
significant and narrow peaks that labeled in Figure~1 and that have 
frequencies (periods) of 
$f_1=0.2421$ cycles~d$^{-1}$ (4.131~d),
$f_2=3.9291$ cycles~d$^{-1}$ (6.108~h),
$f_3=4.1714$ cycles~d$^{-1}$ (5.753~h), and 
$f_4=7.8584$ cycles~d$^{-1}$ (3.054~h).  
We also made periodograms of the light curve from each quarter separately, and 
the frequencies and amplitudes of the peaks nearest these mean frequencies are listed in 
Table~1 and plotted in Figure~3.  The typical measurement uncertainties 
associated with the peak 
frequencies are 0.03 cycle~d$^{-1}$ for quarter 1 and 0.01 cycle~d$^{-1}$
for the subsequent quarters.  We see that $f_1$ and $f_3$ grew significantly 
in strength up to quarter 7 while $f_2$ and $f_4$ remained approximately constant
in amplitude.  

\placetable{tab1}      
 
\begin{figure} 
 {\includegraphics[angle=90,height=12cm]{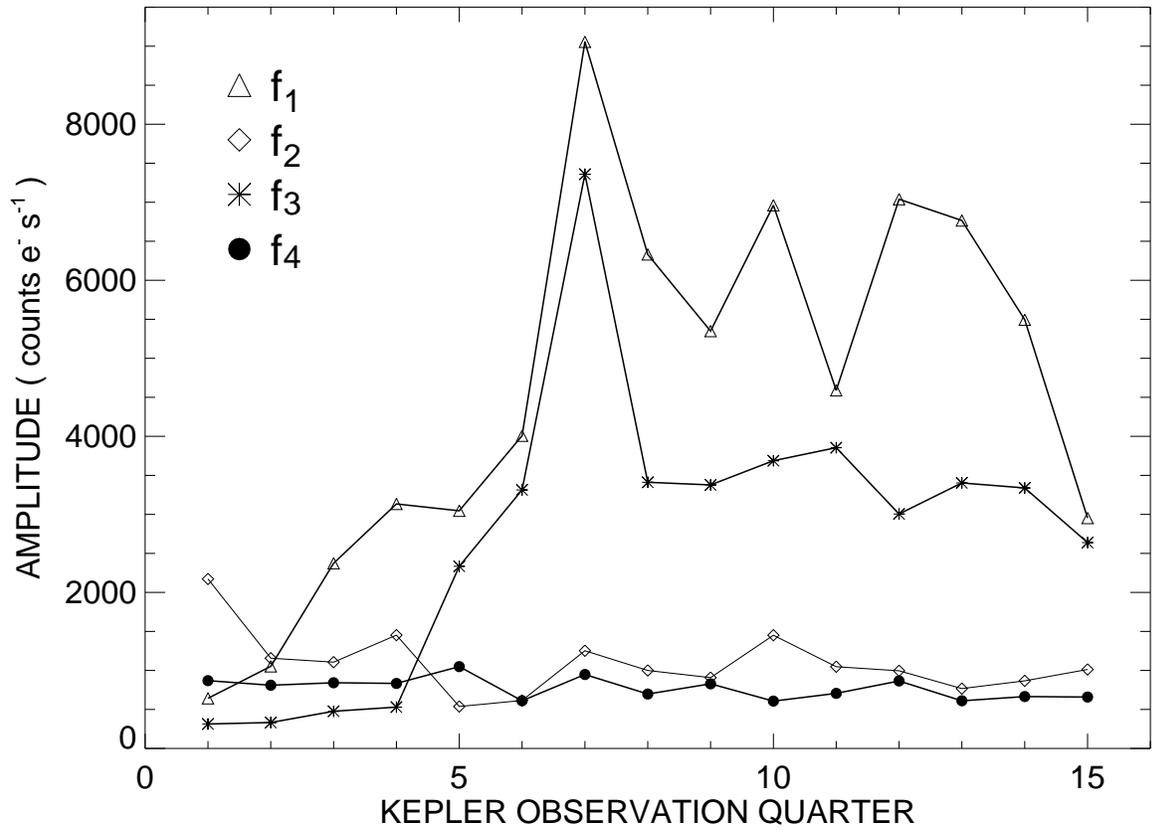}} 
\caption{The amplitudes of one low and three high frequency 
signals in the periodograms of the {\it Kepler} light curves plotted 
against observing quarter number.  The symbols represent  
the amplitudes of the signals at $f_1 = 1/4.131$~d, $f_2=1/6.108$~h, 
$f_3=1/5.753$~h, and $f_4=1/3.054$~h (see Table~1).
\label{fig3}} 
\end{figure}

\placefigure{fig3}     
 
In order to explore this change in the strength of the periodic signals over time, 
we also performed a wavelet analysis that is displayed in the central panel of Figure~1. 
We made the wavelet analysis using the package of Torrence \& Compo 
(1998)\footnote{http://paos.colorado.edu/research/wavelets/}.
The wavelet amplitude of a discrete time series $x_n$ with a sampling time $\delta t$
is given by a convolution of $x_n$ with a wavelet function $\Psi((t^\prime-t)/s)$, 
\begin{displaymath}
W_n(s)=\sum_{n^\prime =0}^{N-1} x_{n^\prime} \Psi^\star
 \left[{{(n^\prime - n)\delta t}\over{s}}\right]
\end{displaymath}
where $s$ is the wavelet scale, $n$ is the index for the time variable, and 
the superscript $\star$ indicates the complex conjugate.  
The wavelet function used in this paper is the Morlet function defined by 
\begin{displaymath}
\Psi(t/s)=\pi^{-\frac{1}{4}} \exp\left(i \omega {{t}\over{s}}\right) 
          \exp\left(-{1\over2}{{t^2}\over{s^2}}\right)
\end{displaymath}
where $t$ is the time difference and $\omega$ is a dimensionless oscillation
frequency multiplier that sets the number of oscillations within the central part of 
the wavelet function.  We adopted $\omega = 10$ which gave better frequency 
resolution than the default value of $\omega = 6$ in the Torrence \& Compo
wavelet package, but at the cost of somewhat worse temporal resolution 
(De Moortel et al.\ 2004).  The wavelet scale sets the test frequency and the 
effective width of the time window, and a grid of scale lengths was set by the 
geometric series  
\begin{displaymath}
s_{j}=s_{0} 2^{j\delta j} , j=0,1,\cdots,J.
\end{displaymath}
We used twice the average time spacing of {\it Kepler} long cadence data for
$s_0$, and adopted $\delta j =0.25$ for a grid of values up to $J=50$. 
The wavelet analysis is done in a similar way to the short-time
Fourier transform (STFT) analysis, in the sense that the signal is
multiplied with a wavelet function, similar to the window function
in the STFT, and the transform is computed separately for different
segments of the time-domain signal.  The width of the window is
changed as the transform is computed for every single spectral component, 
which is probably the most significant characteristic of the wavelet transform.  
Because the wavelet method is a multi-resolution analysis which was
designed to overcome the resolution problem of STFT, it will give 
good time resolution and poor frequency resolution at high
frequencies and bad time resolution and good frequency resolution
at low frequencies.  Furthermore, the edge effects introduced by the finite
limits of the time series become progressively worse at low frequencies
so that the derived wavelet power becomes unreliable within a ``cone of 
influence'' at the boundaries of the time series (Torrence \& Compo 1998). 

The wavelet power for the {\it Kepler} light curve is shown as a grayscale image 
in the central panel of Figure~1 as a function of both time and frequency.  
In the same way as the periodogram, most of the wavelet power occurs in the 
lower frequency part of the diagram, corresponding to the outbursts and dips. 
However, the periodic signals, $f_1$ to $f_4$, are also seen as the dark horizontal 
bands in the wavelet diagram.  We see the same trends as documented in Figure~3, 
namely the near constancy of the signals $f_2=3.9291$ cycles~d$^{-1}$ (6.108~h) and 
$f_4=7.8584$ cycles~d$^{-1}$ (3.054~h) while the other signals 
$f_1=0.2421$ cycles~d$^{-1}$ (4.131~d) and $f_3=4.1714$ cycles~d$^{-1}$ (5.753~h) 
grow from near invisibility to maxima around BJD~2455400. 

These four frequencies are related in two ways. 
First, $f_4$ is the first harmonic of $f_2$ ($f_4 = 2 f_2$) 
indicating that the $f_2$ signal has a non-sinusoidal shape.  
This is seen in Figure~4 (upper panel) that shows the light curve rebinned 
according to phase in the $f_2$ period.  It resembles that of a low amplitude 
ellipsoidal binary light curve with two unequal minima.  On the other hand, 
the shape of the $f_3$ signal (Fig.~4, lower panel) is approximately sinusoidal.  
The second relation is $f_1 = f_3 - f_2$.  Both $f_1$ and $f_3$ share 
the increase in amplitude towards a maximum in quarter 7 (Fig.~3).  
We argue below that the $f_2$ signal is probably the orbital frequency (section 3)
while the $f_1$ signal may correspond to the precessional frequency 
of a tilted accretion disk (section 5). 

\begin{figure} 
 {\includegraphics[angle=90,height=12cm]{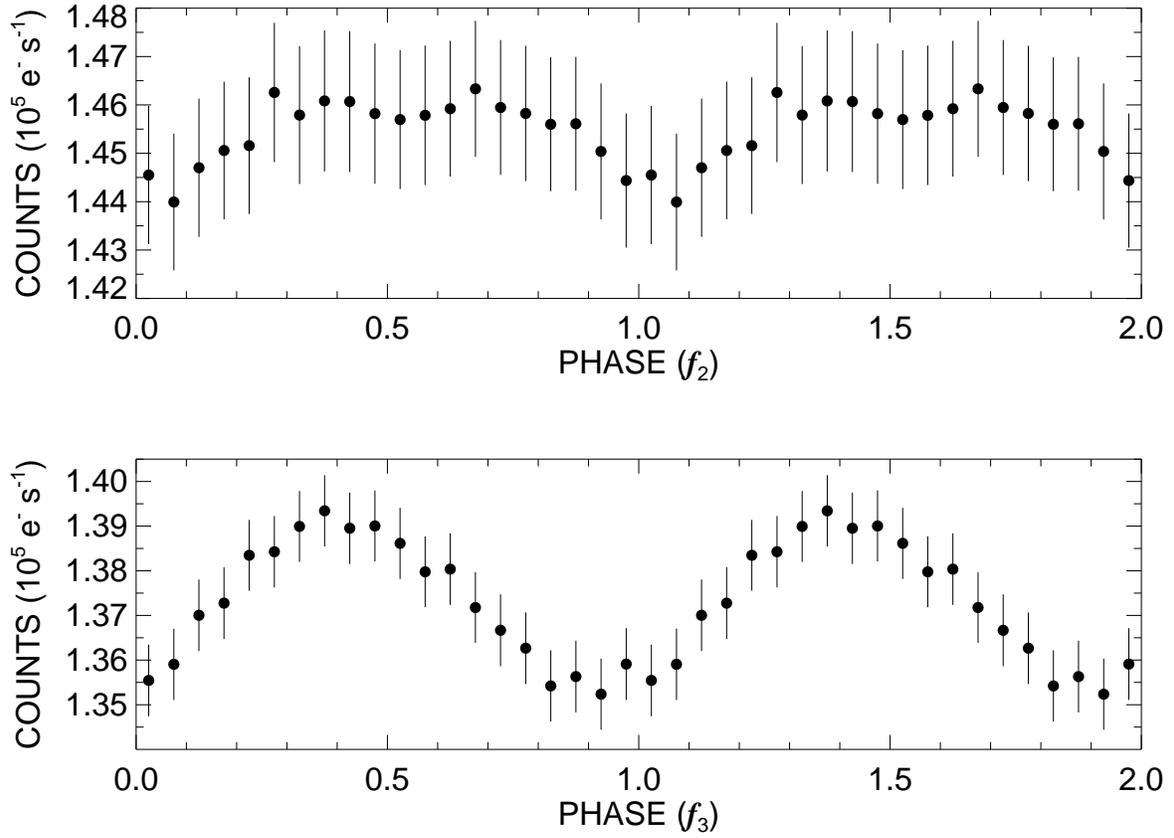}} 
\caption{Phase binned light curves from quarters 1 -- 4 for the 
$f_2$ signal (top panel) and from quarters 5 -- 15 for the $f_3$ signal (lower panel).  
The starting phase is arbitrary in both cases.  Vertical lines indicate the 
square root of the variance of the mean within each bin. 
\label{fig4}} 
\end{figure}

\placefigure{fig4}     

 
\section{Spectroscopy and Orbital Elements}         

We obtained 19 observations of KIC~9406652 with the 
KPNO 4 m Mayall telescope and RC spectrograph with the T2KA CCD detector. 
They were made on three nights over a time span of eight days.
Two lower resolving power ($R=1180$) spectra were made back to back on 
2013 April 14, and these cover most of the optical spectrum (3410 -- 8760 \AA ).
The spectra from 2013 April 20 and 22 have higher resolving power
($R=1940$, KPC 17B grating), and these record the yellow - red portion of the spectrum
(4380 -- 7830 \AA ).  Exposure times ranged from 100 to 600 s.
The observations were reduced and spectra extracted using standard methods
in IRAF\footnote{IRAF is distributed by the National Optical Astronomy Observatory, 
which is operated by the Association of Universities for Research in Astronomy, Inc., 
under cooperative agreement with the National Science Foundation.}  
to create wavelength and flux calibrated spectra.  
These were rectified to a unit continuum by fitting the line-free
regions, and then the spectra were transformed into two matrices on a uniform
$\log \lambda$ wavelength grid (heliocentric frame).  The first of these contains only
the first two spectra that record the blue part of the spectrum,
while the second covers the yellow - red region for all 19 spectra.

Figure 5 shows the blue part of the spectrum
from the average of the first two, lower resolution spectra.  This region
is dominated by broad, hydrogen Balmer lines that are each inscribed with
a central emission line.  The other main features are wide absorption lines
associated with \ion{He}{1} $\lambda\lambda 4026, 4471$, and these also show central
emission.  These are lines often found in spectra of B-type stars, 
but their widths are too large to have a stellar photospheric origin
(unless the spectrum was that of peculiar He-strong star).  The interstellar
lines, such as \ion{Ca}{2} $\lambda 3933$ and the diffuse interstellar band at 4428 \AA , 
are weak or absent, suggesting there is little extinction along the line of
sight to this star.

\begin{figure} 
 {\includegraphics[angle=90,height=12cm]{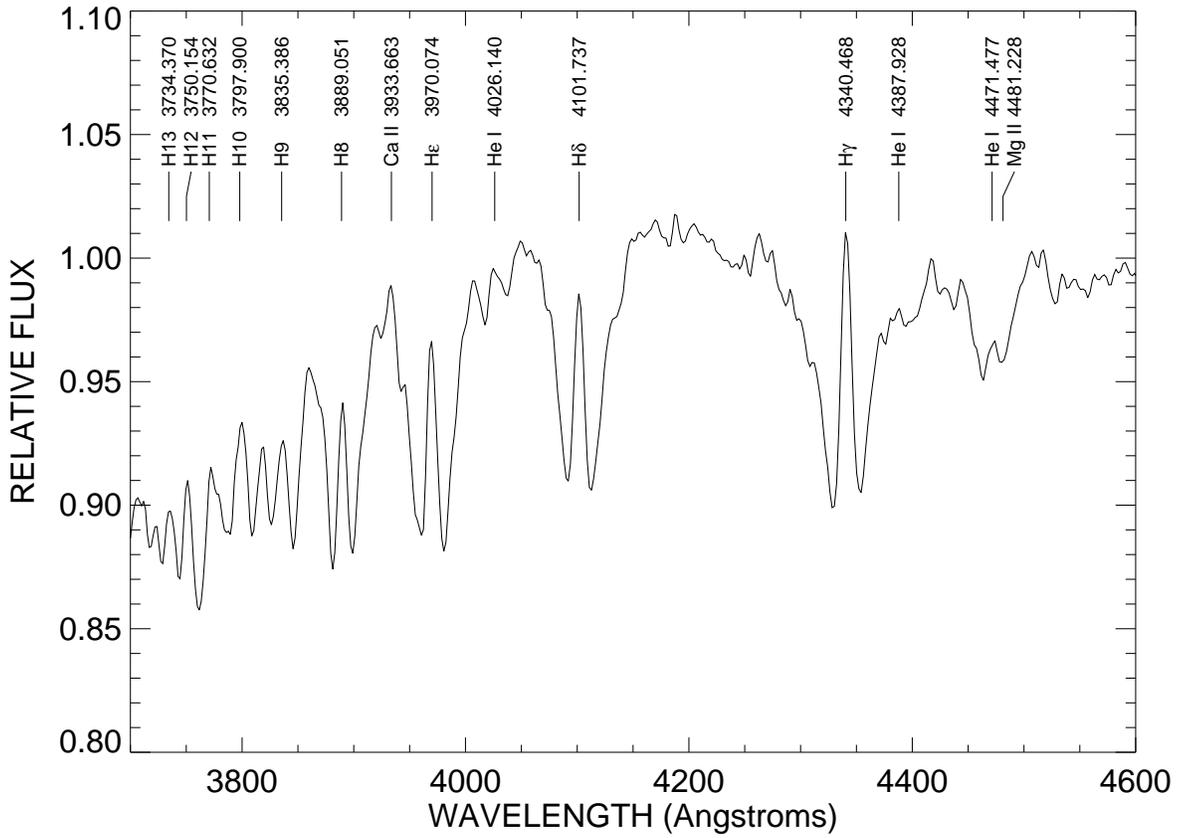}} 
\caption{The average blue spectrum of KIC 9406652 rectified to a 
unit continuum from data collected on the first night.  
Identifications and rest wavelengths of some lines are labeled above. 
\label{fig5}} 
\end{figure}

\placefigure{fig5}     
 
Figure 6 shows the yellow - red spectrum from the average for all 19 spectra. 
The hydrogen Balmer lines of H$\beta$ $\lambda 4861$ and H$\alpha$ $\lambda 6563$ 
both show emission, and the absorption part of the profile is absent in H$\alpha$.
Emission is also present in \ion{He}{1} $\lambda\lambda 5876, 6678, 7065$, with absorption wings
present in \ion{He}{1} $\lambda 5876$.  There are also a number of strong telluric
and molecular bands from Earth's atmosphere that can be identified from the  
atmospheric transmission spectrum shown in the lower part of Figure~6
(from Hinkle et al.\ 2003
\footnote{ftp://ftp.noao.edu/catalogs/atmospheric\_transmission/}).
We checked the wavelength zero-point of each spectrum by cross-correlating
the atmospheric transmission spectrum with each observation (in the topocentric frame). 

\begin{figure} 
 {\includegraphics[angle=90,height=12cm]{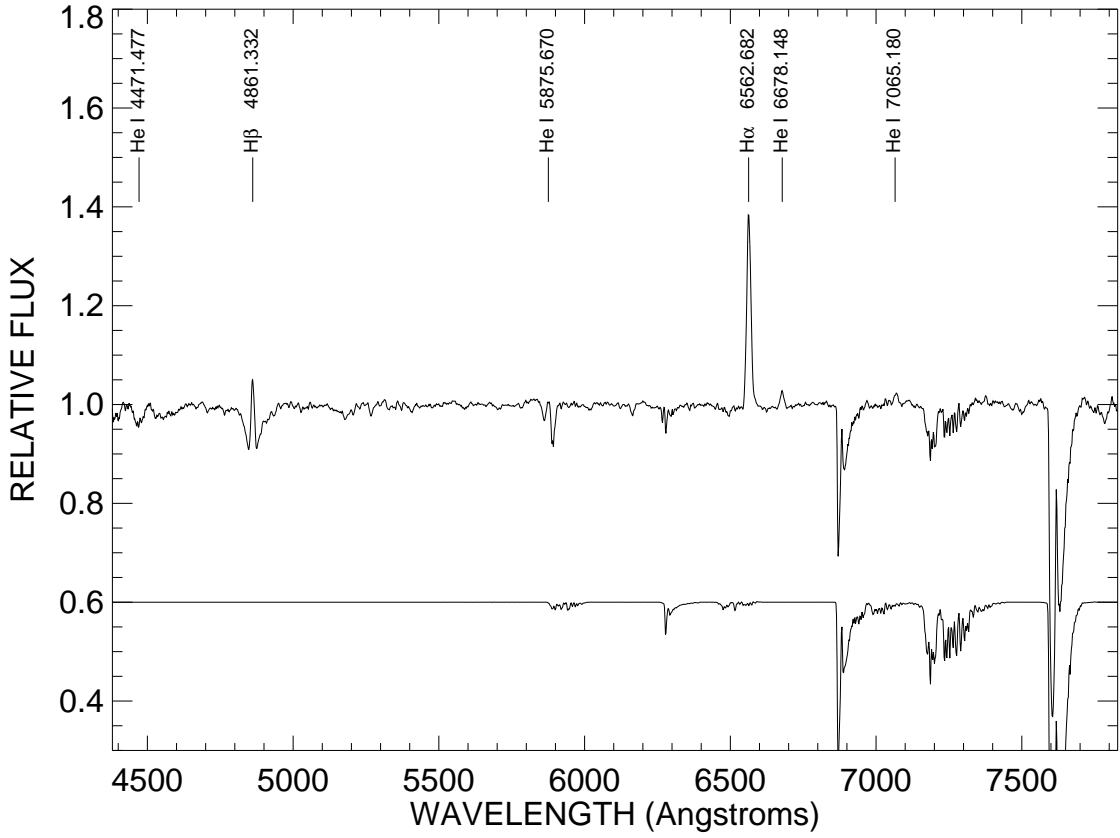}} 
\caption{The average red spectrum of KIC 9406652 rectified to a 
unit continuum from all the data.  The line offset by $-0.4$ shows the 
expected spectral features from Earth's atmosphere (Hinkle et al.\ 2003).
\label{fig6}} 
\end{figure}

\placefigure{fig6}     

The basic appearance of the spectrum is similar to that of the CV 
RW~Sextantis (Beuermann et al.\ 1992), and hence suggestive that KIC~9406652 is also a CV.  
In the spectrum of RW~Sex, the absorption wings form in the accretion disk, and the 
hydrogen Balmer emission lines have two components: a broad base that forms in the 
accretion disk and a narrow peak that originates in the hemisphere of the cool star that faces 
the disk and white dwarf.  All these components display Doppler shifts related to orbital motion.  
Consequently, an important first step is to search for evidence of the orbital modulation of 
the spectral lines of KIC~9406652.

The spectral measurements are summarized in Table~2.  The first
column lists the heliocentric Julian date of mid-observation.  The second column
gives the equivalent width of the strongest emission line, H$\alpha$, which was
measured by numerical integration between 6531 and 6592 \AA .  
Columns 3 -- 9 list various radial velocity measurements of the three strongest
features, H$\alpha$, H$\beta$, and \ion{He}{1} $\lambda 5876$.  The three kinds 
of radial velocity measurements are illustrated for one H$\beta$ profile 
(from HJD 2456404.9342) in Figure~7.  The first type of measurement was a
simple parabolic fit to the upper quarter of the emission peak, and
these are listed under the symbol $V_p$ for each line.  Note that the emission
displayed two close peaks in the last three spectra (for H$\beta$ and the
\ion{He}{1} lines but not H$\alpha$), so there are two entries for $V_p$ in Table 2
for the blue and red peaks.  The spectral evolution of the profiles on the
third night is shown in Figure~8, where each subsequent spectrum
is offset downwards by an amount equal to five times the elapsed time in days.
The other two measurements are bisectors of the line wings from Gaussian sampling
(Shafter et al.\ 1986).  The line wings form in the fastest moving gas, which is 
located close to the white dwarf if the emission forms in an accretion disk.
The columns for $V_w$ give the emission line wing bisector velocities for
H$\alpha$ and H$\beta$ (from positions at $\pm 500$ km~s$^{-1}$), while $V_a$
lists the bisector velocities for the extreme absorption wings of H$\beta$ 
(at $\pm 1400$ km~s$^{-1}$) and \ion{He}{1} $\lambda 5876$ (at $\pm 1300$ km~s$^{-1}$).
All these velocity measurements show similar trends, and we averaged the five
measurements for the emission components (see Table 2, columns 10 and 11 for the mean 
and standard deviation of the mean).  We did not include the absorption wing
velocities because they have a large scatter and possible systematic differences.  
 
\begin{figure} 
 {\includegraphics[angle=90,height=12cm]{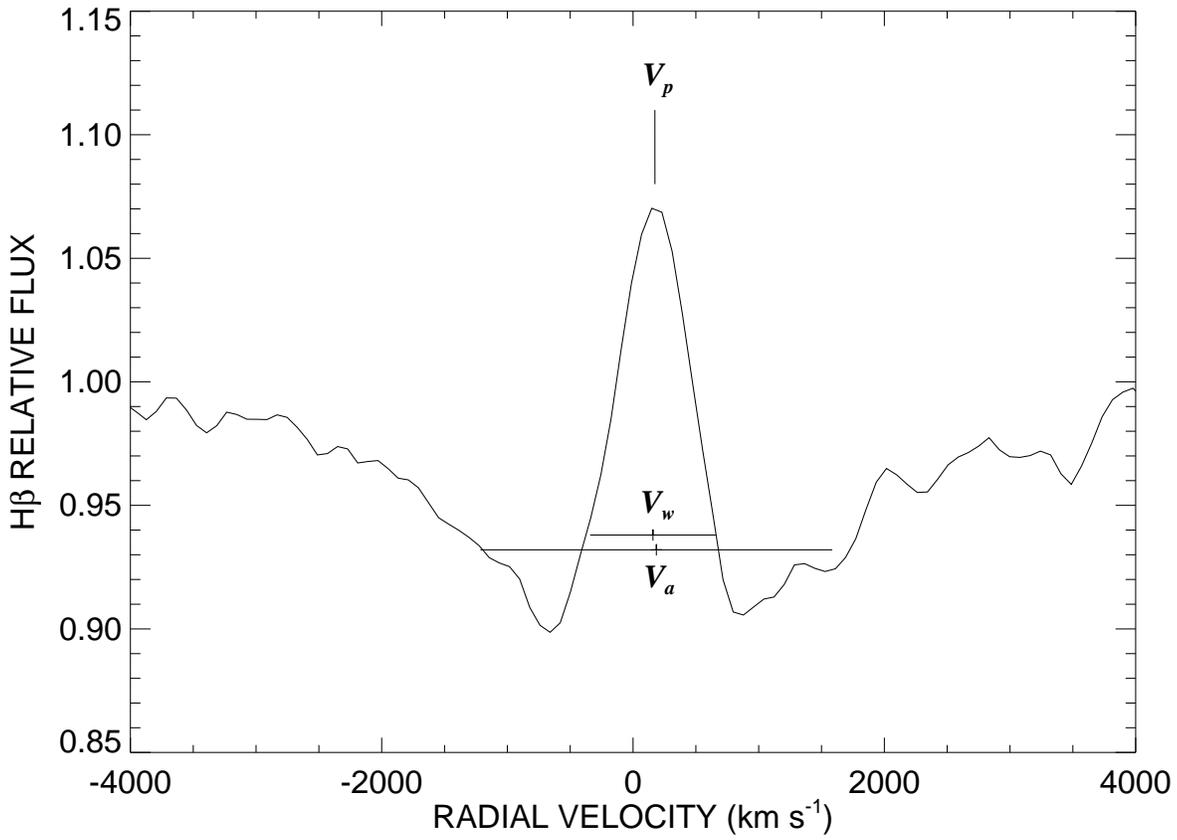}} 
\caption{The radial velocity measurements obtained for the H$\beta$ profile 
(from the spectrum obtained on HJD 2456404.9342): a parabolic fit 
of the upper peak ($V_p$) and Gaussian sampled bisectors of the emission 
wings ($V_w$) and the absorption wings ($V_a$).
\label{fig7}} 
\end{figure}

\placefigure{fig7}     

\begin{figure} 
 {\includegraphics[angle=90,height=12cm]{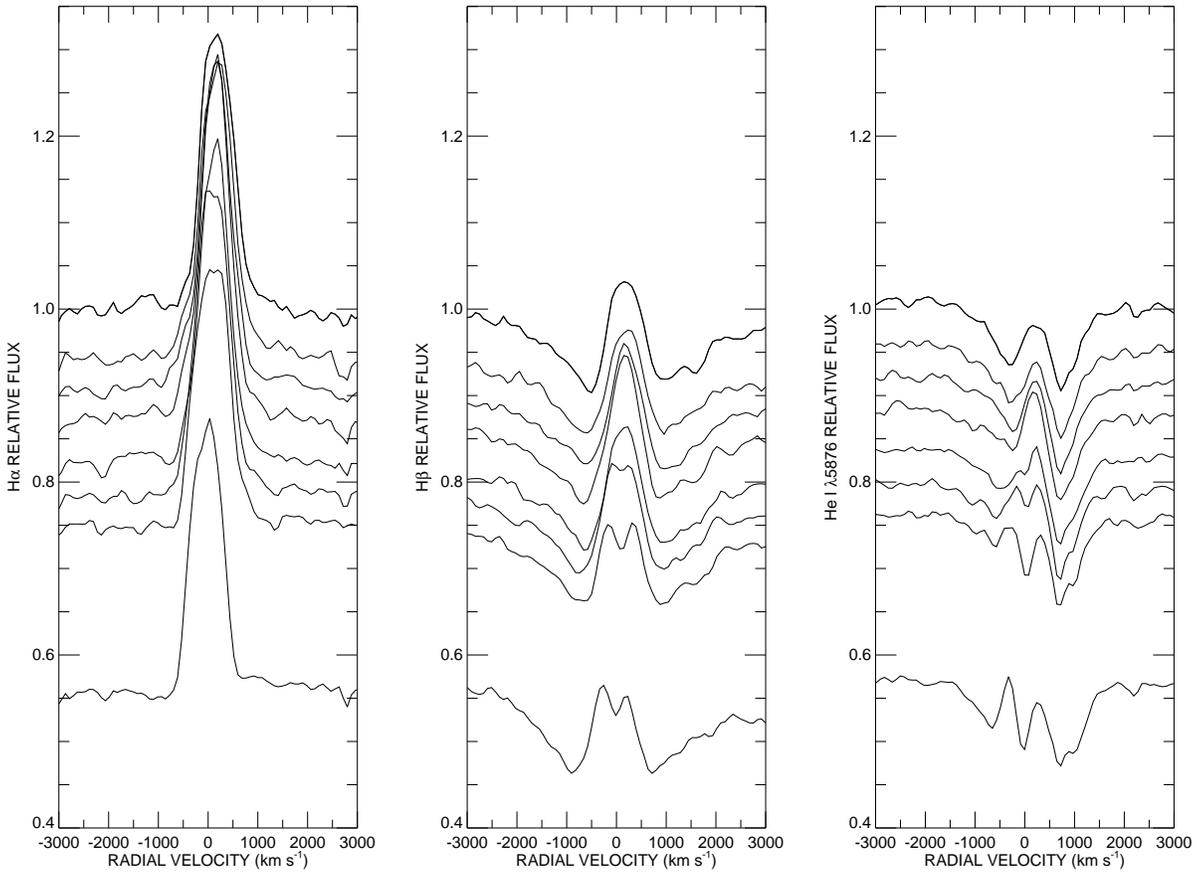}} 
\caption{A montage of the line profiles of H$\alpha$, H$\beta$, and 
\ion{He}{1} $\lambda 5876$ observed on the third night.  The continua 
of each spectrum are offset downwards from the first spectrum (top)
by an amount equal to five times the elapsed time in days. 
\label{fig8}} 
\end{figure}

\placefigure{fig8}     

\placetable{tab2}      
 
If KIC~9406652 is a CV, then we would expect that the companion is a cool star. 
The optical spectra of K- and M-dwarfs are dominated by broad molecular bands
(Gray \& Corbally 2009), but it is difficult to search for such broad features 
in the rectified versions of our spectra because the rectification process 
tends to remove low frequency patterns.  However, we were able to identify 
a significant depression in the average spectrum in the region of 5200 \AA 
~(see Fig.~6) that resembles the MgH and \ion{Mg}{1}~b blend found in cool
star spectra (see Fig.~8.1 of Gray \& Corbally 2009).  We measured the radial 
velocity of this complex through cross-correlation with a model spectrum 
from the BT-Settl PHOENIX grid from Rajpurohit et al.\ (2013)
\footnote{Available from France Allard at http://phoenix.ens-lyon.fr/Grids/BT-Settl/CIFIST2011/SPECTRA/}.
We selected a model with $T_{\rm eff} = 3900$~K, $\log g = 5.0$, and solar
metallicity, parameters appropriate for a M0~V dwarf star (however, this 
choice is not critical because we show below in Fig.~11 that a model for hotter  
$T_{\rm eff} = 4500$~K star shows a very similar spectral morphology and 
would presumably give nearly identical radial velocity measurements).  We then 
re-binned and rectified the model spectrum in the same way as done for the observed
spectra (see Fig.~11 below).  We cross-correlated each observed spectrum with a flux 
diluted version of the model spectrum over the wavelength range between 5013 and 6221 \AA 
~(omitting the region around \ion{He}{1} $\lambda 5876$).  The resulting 
cross-correlation peaks were well-defined in all but the results for first two
low dispersion spectra, and the corresponding cross-correlation function (ccf) 
radial velocities and their uncertainties (Zucker 2003) are listed in columns 
12 and 13 of Table 2, respectively. 
 
Both the emission and absorption feature radial velocities show large excursions
over the course of the second and third nights, so we searched for a periodic 
signal by calculating the discrete Fourier transform of each velocity set 
(Roberts et al.\ 1987).  The resulting periodograms, shown in Figure~9, have a large 
but mutually consistent set of peaks due to the many possible alias frequencies
consistent with our limited spectroscopic time series.  The two similar signal frequencies
from the analysis of the {\it Kepler} light curve (section 2), $f_2$ and $f_3$, 
are also indicated in Figure~9, and only $f_2$ is consistent with the variations
in the radial velocities. 

\begin{figure} 
 {\includegraphics[angle=90,height=12cm]{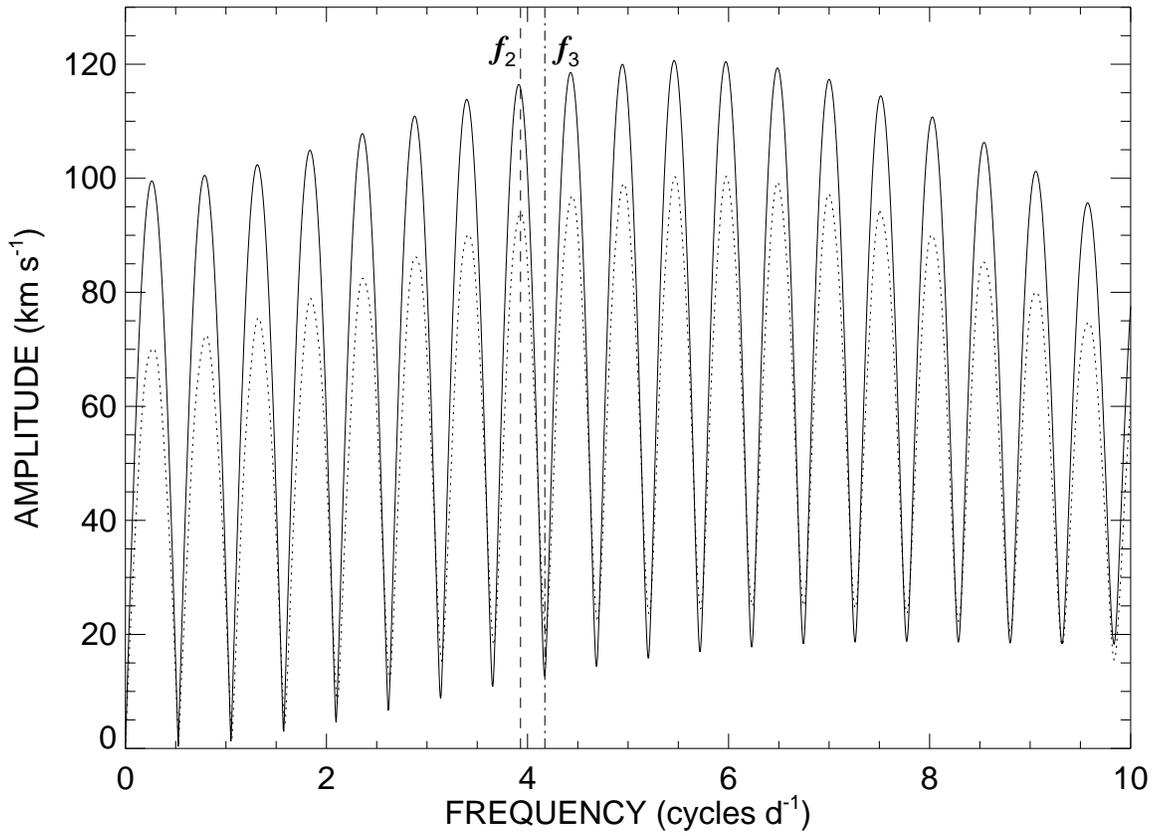}} 
\caption{Periodograms for the radial velocities of the absorption
(solid line) and the emission lines (dotted line) compared 
with the two main signals in the {\it Kepler} light curve
($f_2$ and $f_3$; see section 2 for details).
\label{fig9}} 
\end{figure}

\placefigure{fig9}     
 
We made several estimates of the orbital elements using the 
non-linear, least squares fitting program of Morbey \& Brosterhus (1974), 
and these are summarized in Table~3.  We assumed circular orbits given 
the short orbital period, and we made solutions for both the emission 
and absorption radial velocities with the period free and fixed to 
the mean for $f_2$ from Table~1, $P=0.25440 \pm 0.00008$~d. 
The solutions for the emission line measurements are given under 
the columns labeled Emission, and those for the absorption line ccf velocities 
under columns labeled Absorption.  The rows list the orbital period $P$,
the epoch of cool star inferior conjunction $T$, the semiamplitude $K$, 
the systemic velocity $\gamma$, and the root mean square of the 
residuals from the fit.   The radial velocity curves for the 
period-free solutions are illustrated in Figure~10.  
This plot depicts the mean emission line velocity measurements as 
open circles and the absorption line ccf measurements as filled circles. 
The results are generally consistent between the period free and fixed solutions.   
The systemic velocities for the emission and absorption line systems
are also significantly different, but this is not surprising given the 
differences in the nature of the measurements.  The emission lines, for example, 
may have subtle blue absorption components if a disk wind exists, and this
would tend to yield emission measurements biased to more positive values. 
Furthermore, small differences between the cool star's spectrum and the
adopted model would also produce a net velocity difference from the true
systemic velocity.  Finally, we caution that the hemisphere of the companion 
facing the hot white dwarf may be significantly heated by the flux of the white
dwarf and disk.  This would shift the center of light of the companion 
flux away from the star's center of mass towards the white dwarf, and 
this would cause us to underestimate the true semiamplitude of the 
companion's radial velocity curve.

\placetable{tab3}      
 
\begin{figure} 
 {\includegraphics[angle=90,height=12cm]{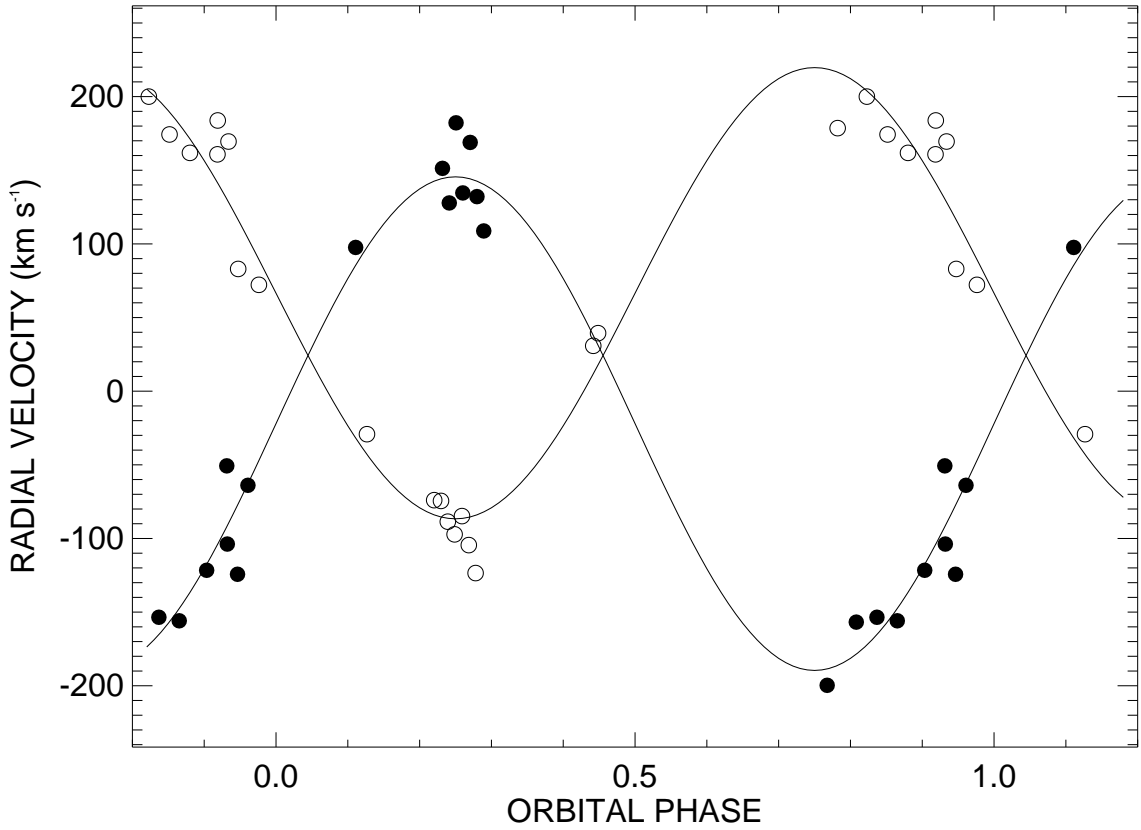}} 
\caption{The preliminary radial velocity curves for the companion star 
(filled circles) and accretion disk (open circles) based upon the independent 
period fitting solutions in Table~3 (columns 4 and 2, respectively).
The measurement uncertainties are approximately 14 km~s$^{-1}$ (see Table 2).
No companion star measurement was feasible for the two low resolution spectra
obtained near orbital phase 0.45.
\label{fig10}} 
\end{figure}

\placefigure{fig10}    

We estimate that minimum light for $f_2$ occurred in the 
quarter 15 {\it Kepler} light curve at BJD~$2456286.646 \pm 0.005$ 
(based upon an $f_2$ phase-binned light curve for that quarter), and the 
epoch of cool star inferior conjunction in our radial velocity data occurred after 
an elapsed time of $118.324 \pm 0.006$~d.  This corresponds to a duration 
of $456.10 \pm 0.14$ orbital cycles, where the large uncertainty is derived 
from the number of cycles and uncertainty in $P$.  This near integer relation 
suggests that minimum light in the $f_2$ cycle occurs at cool star inferior
conjunction. 

 
\section{Spectral Properties of the Components}     

We used a Doppler tomography algorithm (Bagnuolo et al.\ 1994)
to reconstruct the individual spectra of the disk and companion. 
The algorithm performs the spectral reconstructions 
using the calculated radial velocity curves (from 
the solutions where the period was fit) and an estimate of the 
flux ratio $F_2/F_1$, where $F_2$ and $F_1$ are the fluxes 
of the secondary and disk, respectively. 
The method assumes for simplicity that this monochromatic 
flux ratio is constant across the spectrum and at all orbital 
phases, and both of these assumptions are suspect in this case. 
Nevertheless, the reconstruction is acceptable over a limited spectral 
range (in particular omitting those regions with terrestrial 
atmospheric lines that do not share in the orbital motion of 
either component), and we show the reconstructed spectrum of the 
companion in Figure~11.  Also shown are model spectra from 
the grid of Rajpurohit et al.\ (2013) for solar abundance atmospheres
with $T_{\rm eff} =4500$~K and $\log g = 4.5$ (above) and 
with $T_{\rm eff} =3900$~K and $\log g = 5.0$ (below; 
used as the ccf template in section 3). 
Both model spectra were transformed to the observed wavelength grid
using the observed spectral resolution and were rectified to a 
unit continuum in the same way as was done for the observed spectra. 
A good match of the line depths in the reconstructed spectrum 
with those in the models was made adopting a flux ratio of 
$F_2/F_1 = 0.04 \pm 0.01$ (for a central wavelength of 5400 \AA ). 
The agreement between the observed and model spectra is satisfactory 
over this wavelength range, but because the relative line depths change only modestly 
between the temperatures of the models, it is difficult to estimate 
the effective temperature of the companion from the features in this 
part of the spectrum. 

\begin{figure} 
 {\includegraphics[angle=90,height=12cm]{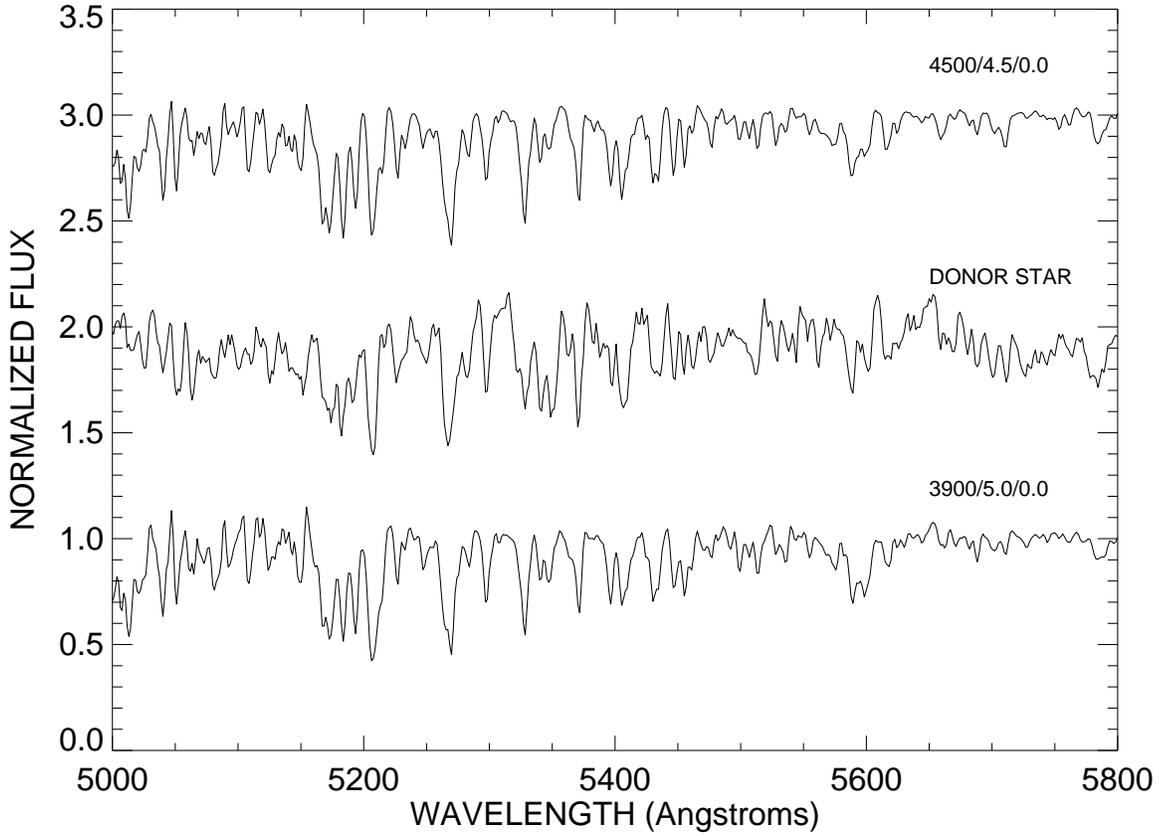}} 
\caption{The reconstructed spectrum of the companion star (middle)
compared with models for $T_{\rm eff} = 4500$ K (above) and 
$T_{\rm eff} = 3900$ K (below). Each model is labeled with the
adopted values of $T_{\rm eff}$ / $\log g$ / $\log A$ (where $A$
is the metal abundance relative to that of the Sun). 
The spectra are offset by 0, +1, and +2 continuum units for clarity. 
\label{fig11}} 
\end{figure}

\placefigure{fig11}    
 
Figure 12 shows the reconstructed spectrum of the hot source 
that follows the emission line radial velocities.  This appears similar
to the average spectrum (Fig.~6) because the most of the flux in 
this region originates in the accretion disk surrounding the white dwarf. 
One interesting difference, however, is seen in the appearance of 
\ion{He}{1} $\lambda 5876$ line, which shows absorption wings that 
are similar in shape to those of H$\beta$.  The stronger red wing of 
\ion{He}{1} $\lambda 5876$ line in the average spectrum is due to 
blending with the strong \ion{Na}{1}~D $\lambda\lambda 5890, 5896$ feature 
in the spectrum of the cool companion.  There are a number of similarities
in the reconstructed spectrum of the disk with that of RW~Sex 
(Beuermann et al.\ 1992; see their Fig.~2) including the presence of emission in the 
\ion{C}{3}/\ion{N}{3} complex near 4650 \AA ~and in \ion{He}{2} $\lambda 4686$,
which probably form in the hotter, inner part of the disk close to the 
white dwarf. 

\begin{figure} 
 {\includegraphics[angle=90,height=12cm]{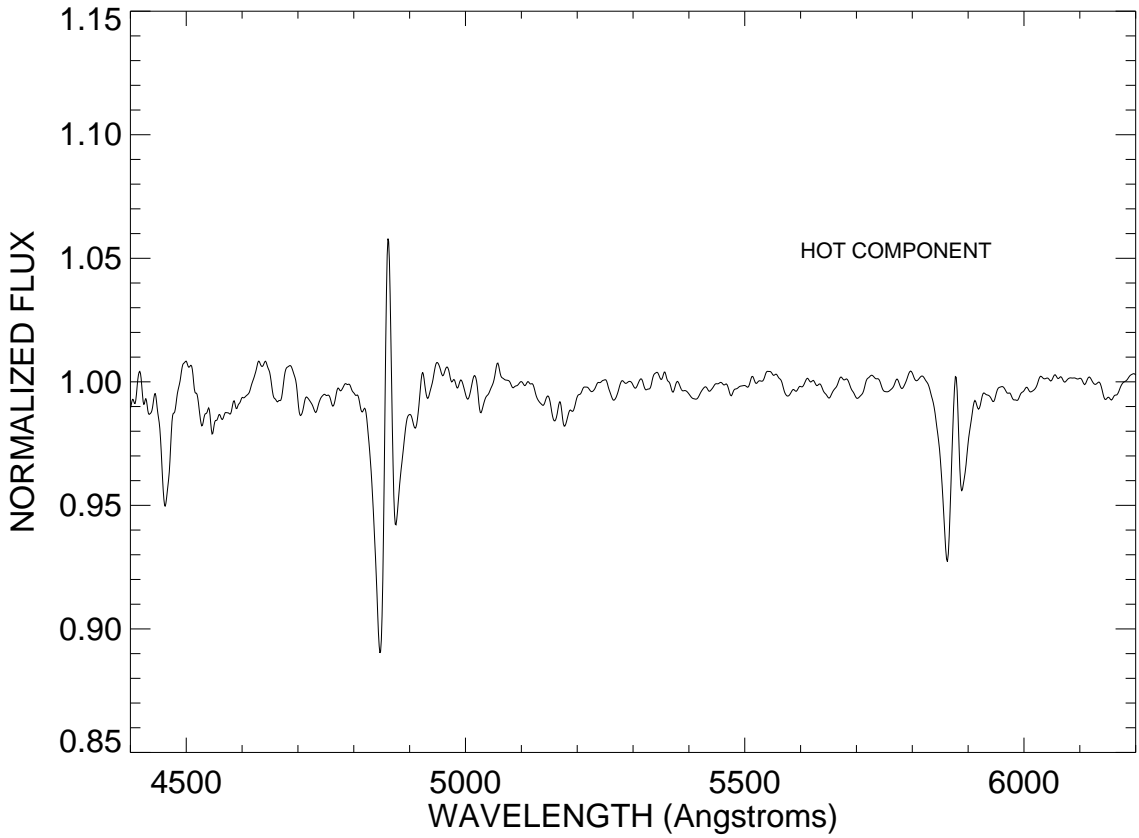}} 
\caption{The reconstructed spectrum of the accretion disk. 
The spectrum was smoothed with a Gaussian transfer function with 
a FWHM = 5 pixels for clarity.  The \ion{He}{1} $\lambda 5876$ 
profile appears similar to that of H$\beta$ here because the 
tomographic reconstruction assigned the blended \ion{Na}{1} doublet 
to the spectrum of the donor (cf.\ Fig.~6).
\label{fig12}} 
\end{figure}

\placefigure{fig12}    

We show a representation of the spectral energy distribution (SED) 
in Figure~12.  The small plus signs show the flux in 200 \AA ~bins from 
one spectrum on each of the three nights of observation, and the spread in 
these is consistent with the amplitude of the short term photometric variability 
and the observational errors in the flux.  The larger plus signs 
represent flux measurements from broad band photometry: 
a GALEX FUV magnitude ($\lambda_{\rm eff}=1516$ \AA ; Morrissey et al.\ 2005),
Johnson $UBV$ photometry (Everett et al.\ 2012),
2MASS $JHK$ photometry (Skrutskie et al.\ 2006), and
WISE 3.35, 4.6, and $11.6 ~\mu$m photometry (Jarrett et al.\ 2011). 
Note that we omitted the WISE $22.1 ~\mu$m measurement because  
no uncertainty estimate was listed.  
Also shown are several model flux distributions that are all modified
for interstellar extinction (Fitzpatrick 1999) assuming a value 
of reddening appropriate for this direction in the Galaxy, 
$E(B-V)= 0.07$~mag (Schlafly \& Finkbeiner 2011; Gontcharov 2012).  
These model flux curves are all normalized to the observed flux 
at 5400 \AA ~(interpolated between $B$ and $V$ from Everett et al.\ 2012)
with each component's contribution set by the observed flux ratio
at that wavelength, $F_2/F_1 = 0.04$.  The dash-dotted line shows a Planck curve 
for $T=17450$~K as a representation of the disk flux contribution, and the 
dotted and dashed curves show companion fluxes (from Rajpurohit et al.\ 2013)
for $T_{\rm eff} = 3900$~K and 4500~K, respectively.  The solid lines show
the sum of the fluxes of the disk and companion, and the model with a 
cooler companion ($T_{\rm eff} = 3900$~K; upper curve) shows a somewhat better
match to the infrared photometry.  We caution, however, that the photometry 
was gathered over a long time span and that intrinsic variability of the 
source will influence the appearance of the SED.  For example, if the 
models were normalized at 5400 \AA ~to the brighter phase recorded in our
spectrophotometry, then the model with a $T_{\rm eff} = 4500$~K companion 
would probably fit the IR photometry as well. 
 
\begin{figure} 
 {\includegraphics[angle=90,height=12cm]{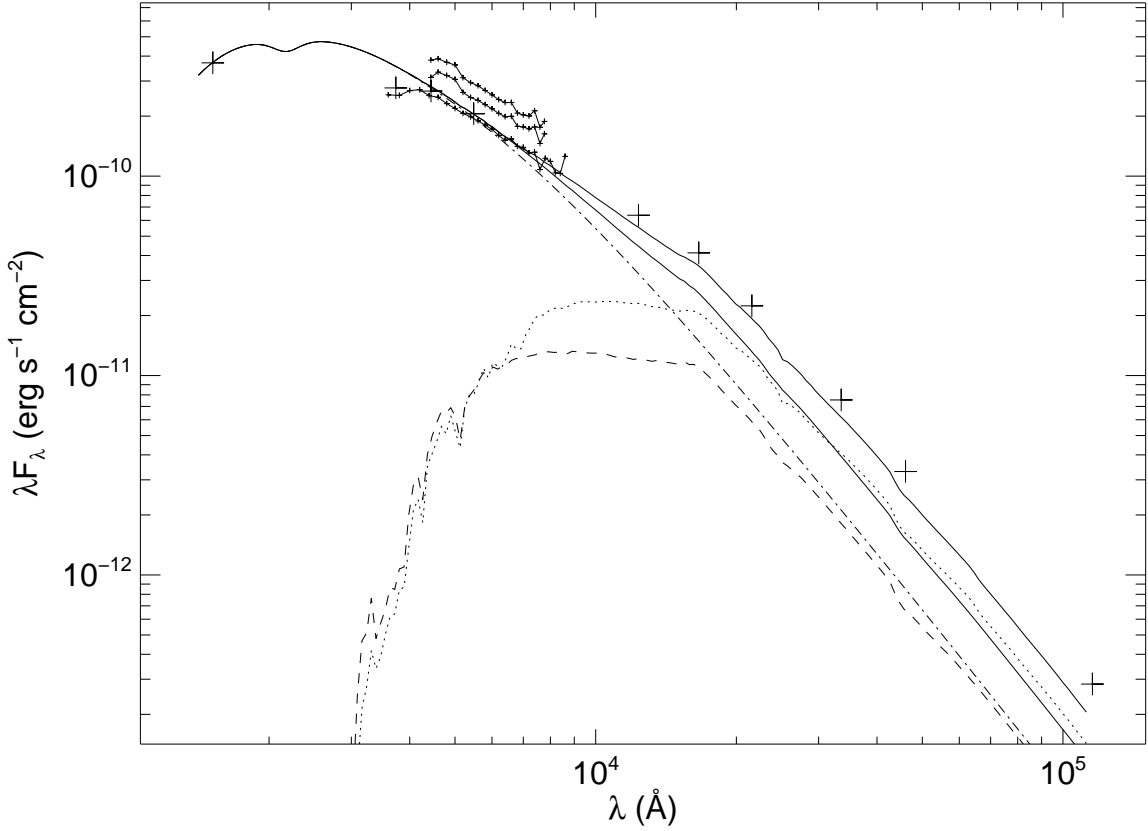}} 
\caption{The spectral energy distribution of KIC 9406652.
The large plus signs indicate multi-wavelength photometric measurements  
(described in the text) while the small plus signs shows our 
spectrophotometric results for each of the three nights. 
The dash-dotted line shows a Planck curve for $T=17450$~K to
represent the disk flux contribution, and the dotted and dashed 
curves show possible companion fluxes for $T_{\rm eff} = 3900$~K and 
4500~K, respectively (normalized to the observed fluxes at 5400 \AA ).
All these are transformed for interstellar extinction assuming $E(B-V)=0.07$ mag.
The upper and lower solid lines show the sum of the Planck and 
companion fluxes for $T_{\rm eff} = 3900$~K and 4500~K, respectively.
\label{fig13}} 
\end{figure}

\placefigure{fig13}    

 
\section{Discussion}                                

The spectra of KIC~9406652 show clearly that the object is a CV. 
We observe the spectral components of both the accretion disk
and cool donor star, and the orbital velocity variations indicate
that the 6.108~h periodicity in the light curve is the 
orbital period of the binary.  The object is apparently not 
an X-ray source (A.\ Smale, private communication), so we expect 
that the mass gainer is a white dwarf as found in other CVs. 
However, the light curve is remarkable for its recurrent outbursts 
that although bright are not as large as those observed in dwarf novae. 
The light properties resemble those of a group of old novae and 
nova-like CVs identified by Honeycutt et al.\ (1998) that 
display unusual ``stunted'' outbursts.  These objects experience
quasi-periodic outbursts that are often accompanied by fadings 
or ``dips'' in the light curve.  One of the objects in this group
is RW~Sex (Beuermann et al.\ 1992), which has a similar orbital period
($P=5.88$~h) and shares a number of spectral similarities with KIC~9406652. 
Thus, the {\it Kepler} light curve of KIC~9406652 provides us with a key opportunity 
to study the processes that lead to  ``stunted'' outbursts in such CVs.

We can use the preliminary orbital elements to estimate the physical properties 
of the binary system.  If we tentatively adopt the fixed $P$ elements
(Table 3), then the mass ratio is $q = M_2 /M_1 = 0.83 \pm  0.07$, 
the projected semimajor axis is $a \sin i = 1.54 \pm 0.06 ~R_\odot$, and 
the mass products are $M_1 \sin^3 i = 0.41 \pm 0.04 ~M_\odot$ and 
$M_2 \sin^3 i = 0.34 \pm 0.04 ~M_\odot$.  The long term evolution of
CVs depends on the properties of the donor star, and given a mass -- radius 
relation for such low mass stars, there exists a relationship between 
the orbital period and donor mass (Patterson 1984; Howell et al.\ 2001; 
Knigge 2006; Knigge et al.\ 2011).  Based upon the semi-empirical relations
with orbital period given by Knigge (2006) and Knigge et al.\ (2011), 
a CV with an orbital period of 6.108~h has a donor of mass $M_2 = 0.75~M_\odot$,
radius $R_2 = 0.72 R_\odot$, an effective temperature $T_{\rm eff} = 4390$~K, 
and $K$-band absolute magnitude $K=4.38$ (assuming that the donor 
is not in a state of advanced evolution).  This effective temperature 
is consistent with the appearance of the donor's spectrum (Fig.~11).
Furthermore, the model evolutionary tracks from Knigge et al.\ (2011)
predict that the white dwarf will be hot ($T_{\rm eff} = 40 - 50$~kK), 
which is consistent with the presence of the \ion{He}{2} $\lambda 4686$ 
emission feature in the spectrum associated with the disk (Fig.~12). 
If we adopt the mass estimate for the donor from the orbital period -- 
mass relationship, then 
the mass of the white dwarf (mass gainer) star is $M_1 \approx 0.9~M_\odot$ and 
the inclination is $i\approx 50^\circ$.  The distance derived from the 
estimated absolute magnitude of the donor and its relative flux contribution 
in the SED at the $K$-band wavelength is in the range of 340~pc (3900~K model)
to 400~pc (4500~K model). 

The orbital period determined from spectroscopy is consistent with 
the $f_2$ signal from the {\it Kepler} light curve (\S2).  
The fact that the orbital phased light curve displays a minimum around 
the time of donor inferior conjunction suggests that the orbital part of 
the light curve variations is associated with a reflection effect, i.e., 
the hemisphere of the donor facing the white dwarf appears brighter. 
The stronger $f_3$ signal indicates the presence of a periodic variation 
that is somewhat shorter than the orbital period.  Such near orbital 
period variations are known in many CVs through the presence of 
``superhumps'' in the light curve, and in some cases they appear 
at shorter periods (see the case of TV Col; Retter et al.\ 2003).
Wood \& Burke (2007) argue that these ``negative superhumps'' are 
caused by variations in the flux from the hot spot where 
the mass transfer stream strikes the accretion disk.   They show 
how the orientation of a tilted and precessing accretion disk in a CV 
results in a hot spot location that changes through the orbital 
and precessional cycle.  When the mass stream encounters the nodal line of the disk, 
the hot spot occurs at a relatively large radial distance from the white dwarf.  
Then, as the companion progresses around the orbit, the stream will travel 
over the equatorial plane to arrive at a position closer to the white dwarf 
where the gas is denser, so that the hot spot flux increases.  However, 
instead of seeing two maxima per orbit if the disk were transparent, the 
final result is that an external observer outside of the disk plane will 
witness one variation per orbit because the hot spot will occur below 
the optically thick disk in the second half of the cycle.   The ``negative
superhump'' objects have a shorter superhump period because the disk 
precession is retrograde to the orbit, so that the line of the nodes is 
seen earlier each orbit.  Thus, in this model, the difference between 
the superhump and orbital frequencies is equal to the disk precessional frequency. 

We suggest that this retrograde precessing disk model applies 
to the case of KIC~9406652.  The difference frequency, 
superhump $f_3$ minus orbital $f_2$, is also observed in the
periodogram as $f_1$ which would correspond to the precessional frequency. 
This signal with a period of 4.13~d is clearly seen in the light curve 
at various times (see Fig.~2) and it probably results from changes in 
the projected size of the disk with the precessional cycle.   Thus, the 
higher frequency signals in the light curve appear to be related to two 
primary ``clocks'', the binary orbital and disk precessional periods. 
Larwood (1998) showed that the ratio between the orbital period and 
forced precession period is 
$${{P}\over{P_p}}={3\over 7} {{\mu}\over {(1 + \mu)^{1/2}}} 
  ~\beta^{3/2} R^{3/2} \cos\delta$$ 
where $\mu=M_2/M_1$ is the mass ratio, $\beta$ is the ratio of disk outer radius 
to Roche radius of the white dwarf, $R$ is the ratio of white dwarf Roche radius
to the binary semimajor axis, and $\delta$ is the disk inclination angle 
relative to the orbital plane.  If we assume $\mu=0.83$, $\beta=1$, 
$R=0.40$ (from the mass ratio and the formula from Eggleton 1983), and
$\delta = 10^\circ$ (a representative small disk tilt), 
then the predicted ratio of $P/P_p = 0.064$ is the same within uncertainties 
as the observed ratio of $P/P_p = 0.062$.  

The lower frequency signals in the periodogram of the light curve are 
related to the outbursts and dips that occur on a $\sim 30$~d timescale (Fig.~1). 
These generally take the form of an outburst followed immediately by a dip.
One exception was observed near BJD~2455190 where a strong dip occurred before
an outburst.  Curiously, the next outburst that occurred near BJD~2455240
was also exceptional in its duration.  Both of these events happened just
prior to the appearance of the $f_1$ and $f_3$ signals in the periodogram 
(see Fig.~1), which marks the beginning of the disk precession phase in 
these observations. 

The cause of the stunted outbursts in this system is unknown, but they may be 
related to cycles of changing disk mass.  Mennickent et al.\ (2003) have
discovered a class of Double Periodic Variables among binaries containing 
massive stars in interacting binaries.  They argue that the mass gainers have
reached critical rotation and can no longer easily accrete additional mass
from the donor.  Consequently, the transferred mass builds up in a thick 
accretion torus surrounding the companion, until it is finally released 
into an expanding circumbinary disk.  The cycle of growth and dissipation of 
the disk gas is observed as the longer periodicity in the light curves of
these systems.  It is possible that a similar process is occurring in CVs like 
KIC~9406652.  Models suggest that the mass transfer rate is relatively large
in longer period CVs (Knigge et al.\ 2011).  
The incoming gas from the donor may become trapped in the disk because direct 
accretion is inhibited by the rapid rotation and/or magnetic field of the white dwarf gainer. 
If this is the case with respect to KIC~9406652, 
then the outburst and dip would be related to the growth in physical size and 
subsequent ejection of disk gas from the system (possibly causing some obscuration). 
We speculate that such gas ejection events may provide the torque required 
to cause a disk warp and promote disk precession (as occurs, for example, by the 
action of a disk coronal wind in the X-ray binary Her~X-1; Schandl \& Meyer 1994).

The remarkable richness of the {\it Kepler} light curve of KIC~9406652 provides
us with the opportunity to explore disk physical processes with unprecedented 
temporal coverage.  This is especially important for this CV where the short 
and long period variations appear to be related in a fundamental way. 
If the outbursts are caused by relatively low energy mass ejections, then 
the system may be surrounded by a circumbinary disk of large dimension, 
and high angular resolution observations may provide the means to detect 
and map such an outflow. 

 
\acknowledgments 
 
We thank all the participants of the Planet Hunters program for their
efforts in finding objects like this one.  We also express our thanks 
to Kimberly Sokal and Phil Massey for helping obtain the first spectra and 
to Sergio Dieterich and Richard Wade for their advice with the analysis.  
This paper includes data collected by the Kepler mission. 
Funding for the Kepler mission is provided by
the NASA Science Mission Directorate.
Some of the data presented in this paper were obtained from 
the Mikulski Archive for Space Telescopes (MAST). 
STScI is operated by the Association of Universities for 
Research in Astronomy, Inc., under NASA contract NAS5-26555. 
Support for MAST for non-HST data is provided by the NASA Office of 
Space Science via grant NNX09AF08G and by other grants and contracts.
Our work was supported in part by NASA grant NNX12AC81G (DRG) and 
by the National Science Foundation under grant AST-1009080 (DRG). 
Institutional support has been provided from the GSU College 
of Arts and Sciences and the Research Program Enhancement 
fund of the Board of Regents of the University System of Georgia, 
administered through the GSU Office of the Vice President 
for Research and Economic Development.  

{\it Facilities:} \facility{Kepler, Mayall} 
 


\clearpage
 
 
\begin{deluxetable}{cccccccccc}
\tabletypesize{\scriptsize} 
\rotate 
\tablewidth{0pc} 
\tablenum{1} 
\tablecaption{Light Curve Periodogram Signal Frequencies and Amplitudes\label{tab1}} 
\tablehead{ 
\colhead{Quarter}          & 
\colhead{Mean Date}        & 
\colhead{$f_1$}            & 
\colhead{$a_1$}            & 
\colhead{$f_2$}            & 
\colhead{$a_2$}            & 
\colhead{$f_3$}            & 
\colhead{$a_3$}            &  
\colhead{$f_4$}            & 
\colhead{$a_4$}           \\  
\colhead{Number}            & 
\colhead{(BJD$-$2,400,000)} & 
\colhead{(cycles d$^{-1}$)} & 
\colhead{(e$^-$ s$^{-1}$)}  & 
\colhead{(cycles d$^{-1}$)} & 
\colhead{(e$^-$ s$^{-1}$)}  & 
\colhead{(cycles d$^{-1}$)} & 
\colhead{(e$^-$ s$^{-1}$)}  & 
\colhead{(cycles d$^{-1}$)} & 
\colhead{(e$^-$ s$^{-1}$)}               
} 
\startdata 
\phn 1 & 54981 & 0.217&\phn643& 3.933 &     2172 & 4.174 & \phn 313 & 7.857 & \phn 869 \\      
\phn 2 & 55047 & 0.242 & 1058 & 3.936 &     1157 & 4.195 & \phn 332 & 7.859 & \phn 810 \\
\phn 3 & 55138 & 0.242 & 2374 & 3.929 &     1106 & 4.168 & \phn 477 & 7.858 & \phn 842 \\
\phn 4 & 55230 & 0.242 & 3134 & 3.945 &     1453 & 4.172 & \phn 530 & 7.858 & \phn 833 \\
\phn 5 & 55324 & 0.239 & 3046 & 3.926 & \phn 538 & 4.167 &     2335 & 7.860 &     1049 \\
\phn 6 & 55417 & 0.245 & 4008 & 3.930 & \phn 615 & 4.174 &     3313 & 7.859 & \phn 611 \\ 
\phn 7 & 55508 & 0.241 & 9056 & 3.929 &     1252 & 4.171 &     7359 & 8.097 & \phn 948 \\
\phn 8 & 55602 & 0.242 & 6331 & 3.930 & \phn 999 & 4.169 &     3412 & 7.858 & \phn 697 \\
\phn 9 & 55690 & 0.239 & 5349 & 3.930 & \phn 908 & 4.167 &     3377 & 7.858 & \phn 828 \\
    10 & 55787 & 0.245 & 6961 & 3.930 &     1451 & 4.171 &     3688 & 7.857 & \phn 607 \\
    11 & 55883 & 0.241 & 4589 & 3.928 &     1047 & 4.170 &     3855 & 7.858 & \phn 706 \\     
    12 & 55974 & 0.241 & 7039 & 3.930 & \phn 995 & 4.172 &     3005 & 7.857 & \phn 865 \\   
    13 & 56061 & 0.244 & 6766 & 3.929 & \phn 767 & 4.172 &     3403 & 7.856 & \phn 611 \\      
    14 & 56156 & 0.251 & 5495 & 3.927 & \phn 867 & 4.178 &     3338 & 7.860 & \phn 665 \\      
    15 & 56256 & 0.242 & 2953 & 3.930 &     1012 & 4.170 &     2638 & 7.859 & \phn 659 \\  
\enddata 
\end{deluxetable}

\begin{deluxetable}{ccccccccccccc}
\rotate 
\tabletypesize{\scriptsize} 
\tablewidth{0pc} 
\tablenum{2} 
\tablecaption{Spectroscopic Measurements\label{tab2}} 
\tablehead{ 
\colhead{HJD}              & 
\colhead{$-W_\lambda$}     & 
\colhead{$V_p$}            & 
\colhead{$V_w$}            & 
\colhead{$V_p$}            & 
\colhead{$V_w$}            & 
\colhead{$V_a$}            & 
\colhead{$V_p$}            & 
\colhead{$V_a$}            & 
\colhead{$V_e$}            & 
\colhead{$\sigma(V_e)$}    & 
\colhead{$V_{\rm ccf}$}    & 
\colhead{$\sigma(V_{\rm ccf})$} \\  
\colhead{($-$2,456,000)}   & 
\colhead{(H$\alpha$)}      & 
\colhead{(H$\alpha$)}      & 
\colhead{(H$\alpha$)}      & 
\colhead{(H$\beta$)}       & 
\colhead{(H$\beta$)}       & 
\colhead{(H$\beta$)}       & 
\colhead{(He I)}           & 
\colhead{(He I)}           & 
\colhead{}                 & 
\colhead{}                 & 
\colhead{}                 & 
\colhead{}                 \\
\colhead{(d)}              & 
\colhead{(\AA )}           & 
\colhead{(km s$^{-1}$)}    & 
\colhead{(km s$^{-1}$)}    & 
\colhead{(km s$^{-1}$)}    & 
\colhead{(km s$^{-1}$)}    & 
\colhead{(km s$^{-1}$)}    & 
\colhead{(km s$^{-1}$)}    & 
\colhead{(km s$^{-1}$)}    & 
\colhead{(km s$^{-1}$)}    & 
\colhead{(km s$^{-1}$)}    & 
\colhead{(km s$^{-1}$)}    & 
\colhead{(km s$^{-1}$)}    \\
\colhead{(1)}              & 
\colhead{(2)}              & 
\colhead{(3)}              & 
\colhead{(4)}              & 
\colhead{(5)}              & 
\colhead{(6)}              & 
\colhead{(7)}              & 
\colhead{(8)}              & 
\colhead{(9)}              & 
\colhead{(10)}             &              
\colhead{(11)}             &              
\colhead{(12)}             &              
\colhead{(13)}                           
} 
\startdata 
 396.9768 &   5.12 &\phn\phs18 &\phn\phs 51 &\phn\phn\phs3&\phn\phs 24 &\phn\phs 92 &\phn\phs 58 & 228 &\phn\phs 31 &    10 &\nodata    &\nodata\\
 396.9785 &   5.98 &\phn\phs14 &\phn\phs 55 &\phn\phs  20 &\phn\phs 20 &\phs    202 &\phn\phs 88 & 255 &\phn\phs 39 &    14 &\nodata    &\nodata\\
 402.9191 &   6.36 &\phs   232 &\phs    162 &\phs     169 &\phs    126 &\phn\phs 54 &\phs    230 & 201 &\phs    184 &    21 &\phn $-$51 &    37 \\
 402.9229 &   6.28 &\phs   239 &\phs    102 &\phs     199 &\phn\phs 72 &\phs    115 &\phs    236 & 148 &\phs    170 &    35 &    $-$124 &    26 \\
 402.9954 &   7.37 &\phn $-$51 &\phn  $-$43 &\phn   $-$86 &\phn  $-$75 &     $-$283 &     $-$116 & 195 &\phn  $-$74 &    13 &\phs   151 &    19 \\
 402.9978 &   7.41 &\phn $-$47 &\phn  $-$30 &\phn   $-$93 &     $-$102 &     $-$265 &     $-$100 & 213 &\phn  $-$75 &    15 &\phs   128 &    14 \\
 403.0002 &   7.20 &\phn $-$45 &\phn  $-$45 &      $-$123 &\phn  $-$94 &     $-$306 &     $-$123 & 194 &\phn  $-$89 &    19 &\phs   182 &    15 \\
 403.0026 &   7.69 &\phn $-$54 &\phn  $-$81 &      $-$129 &     $-$133 &     $-$347 &\phn  $-$90 & 211 &\phn  $-$97 &    15 &\phs   135 &    13 \\
 403.0052 &   7.30 &\phn $-$49 &\phn  $-$63 &      $-$101 &     $-$130 &     $-$130 &\phn  $-$80 & 142 &\phn  $-$85 &    14 &\phs   169 &    17 \\
 403.0076 &   7.60 &\phn $-$81 &\phn  $-$69 &      $-$132 &     $-$141 &     $-$213 &     $-$100 & 152 &     $-$105 &    14 &\phs   132 &    17 \\
 403.0100 &   7.75 &\phn $-$85 &\phn  $-$72 &      $-$171 &     $-$172 &\phn\phs 65 &     $-$118 & 268 &     $-$124 &    21 &\phs   109 &    20 \\
 404.9094 &   5.80 &\phs   144 &\phs    190 &\phs     156 &\phs    201 &\phs    302 &\phs    202 & 249 &\phs    179 &    12 &    $-$200 &    14 \\
 404.9197 &   5.94 &\phs   206 &\phs    177 &\phs     204 &\phs    195 &\phs    237 &\phs    218 & 278 &\phs    200 &\phn 7 &    $-$157 &    12 \\
 404.9270 &   6.33 &\phs   152 &\phs    152 &\phs     167 &\phs    186 &\phs    286 &\phs    206 & 349 &\phs    174 &    10 &    $-$154 &    12 \\
 404.9342 &   6.73 &\phs   154 &\phs    143 &\phs     174 &\phs    158 &\phs    186 &\phs    181 & 273 &\phs    162 &\phn 7 &    $-$156 &    12 \\
 404.9439 &   5.88 &\phs   144 &\phs    135 &\phs     144 &\phs    156 &\phs    251 &\phs    226 & 307 &\phs    161 &    17 &    $-$122 &    12 \\
 404.9512 &   6.19 &\phn\phs74 &\phn\phs 92 &\phn   $-$41 &\phs    103 &\phs    169 &     $-$161 & 221 &\phn\phs 83 &    11 &    $-$104 &    14 \\
 404.9512 &\nodata &\nodata    &\nodata     &\phs     241 &\nodata     &\nodata     &\phs    253 & 221 &\nodata     &\nodata&\nodata    &\nodata\\
 404.9585 &   5.36 &\phn\phs96 &\phn\phs 86 &      $-$155 &\phn\phs 91 &\phs    205 &     $-$326 & 275 &\phn\phs 72 &    19 &\phn $-$64 &    13 \\
 404.9585 &\nodata &\nodata    &\nodata     &\phs     336 &\nodata     &\nodata     &\phs    319 & 275 &\nodata     &\nodata&\nodata    &\nodata\\
 404.9966 &   4.79 &\phn\phn$-$5&\phn $-$20 &      $-$262 &\phn  $-$65 &\phn\phs 82 &     $-$320 & 201 &\phn  $-$29 &    10 &\phn\phs98 &    16 \\
 404.9966 &\nodata &\nodata    &\nodata     &\phs     192 &\nodata     &\nodata     &\phs    278 & 201 &\nodata     &\nodata&\nodata    &\nodata\\
\enddata 
\end{deluxetable}
  
\newpage 
 
\begin{deluxetable}{lcccc} 
\tabletypesize{\small} 
\tablewidth{0pc} 
\tablenum{3} 
\tablecaption{Preliminary Orbital Elements\label{tab3}} 
\tablehead{ 
\colhead{Element}    & 
\colhead{Emission}   &
\colhead{Emission}   &
\colhead{Absorption} & 
\colhead{Absorption} 
}
\startdata 
$P$~(d)               \dotfill & $0.2531 \pm 0.0006$ & 0.2544\tablenotemark{a} 
                               & $0.2540 \pm 0.0012$ & 0.2544\tablenotemark{a} \\ 
$T$ (HJD$-$2,456,000) \dotfill & $404.965\pm 0.006$  & $404.987\pm 0.004$  
                               & $404.968\pm 0.005$  & $404.970\pm 0.004$      \\ 
$K$ (km s$^{-1}$)     \dotfill & $153 \pm 10$        & $139 \pm  8$    
                               & $168 \pm 10$        & $167 \pm 10$            \\ 
$\gamma$ (km s$^{-1}$)\dotfill & $ 66 \pm 11$        & $ 34 \pm 8$    
                               & $-22 \pm 11$        & $-21 \pm 9$             \\ 
rms (km s$^{-1}$)     \dotfill & 26                  & 30           
                               & 26                  & 25                      \\ 
\enddata 
\tablenotetext{a}{Fixed.}
\end{deluxetable} 

\clearpage



\end{document}